\documentclass[12pt]{article}
\usepackage{graphicx}
\usepackage{color}
\usepackage{cite}

\def \beq{\begin{equation}}
\def \eeq{\end{equation}}
\def\eqref#1{(\ref{#1})}
\def\bea{\begin{eqnarray}}
\def\eea{\end{eqnarray}}

\def\URLtilde{\lower0.2em\hbox{$\tilde{\phantom{a}}$}}
\def\mycomm#1{\hfill\break\strut\kern-3em{\color{red}\tt ====> #1
\color{black}}\hfill\break}

\newcount\timecount
\newcount\hours \newcount\minutes  \newcount\temp \newcount\pmhours
\hours = \time
\divide\hours by 60
\temp = \hours
\multiply\temp by 60
\minutes = \time
\advance\minutes by -\temp
\def\hour{\the\hours}
\def\minute{\ifnum\minutes<10 0\the\minutes
\else\the\minutes\fi}
\def\clock{
\ifnum\hours=0 12:\minute\ AM
\else\ifnum\hours<12 \hour:\minute\ AM
\else\ifnum\hours=12 12:\minute\ PM
\else\ifnum\hours>12
\pmhours=\hours
\advance\pmhours by -12
\the\pmhours:\minute\ PM
\fi
\fi
\fi
\fi
}

\def\monthname{\relax\ifcase\month 0/\or January\or February\or
March\or April\or May\or June\or July\or August\or September\or
October\or November\or December\else\number\month/\fi}

\def\bold#1{\setbox0=\hbox{$#1$}     \kern-.025em\copy0\kern-\wd0
\kern.05em\copy0\kern-\wd0
\kern-.025em\raise.0433em\box0 }

\begin{document}
\setcounter{footnote}{1}
\rightline{EFI 17-14}
\rightline{TAUP 3019/17}
\rightline{arXiv:1706.06961}
\vskip1.5cm

\centerline{\large \bf Isospin splittings in baryons with two heavy quarks}
\bigskip

\centerline{Marek Karliner$^a$\footnote{{\tt marek@proton.tau.ac.il}}
 and Jonathan L. Rosner$^b$\footnote{{\tt rosner@hep.uchicago.edu}}}
\medskip

\centerline{$^a$ {\it School of Physics and Astronomy}}
\centerline{\it Raymond and Beverly Sackler Faculty of Exact Sciences}
\centerline{\it Tel Aviv University, Tel Aviv 69978, Israel}
\medskip

\centerline{$^b$ {\it Enrico Fermi Institute and Department of Physics}}
\centerline{\it University of Chicago, 5620 S. Ellis Avenue, Chicago, IL
60637, USA}
\bigskip
\strut

\begin{quote}
\begin{center}
ABSTRACT
\end{center}
Isospin splittings in baryons with two heavy quarks and a $u$ or $d$ quark
are calculated using simple methods proposed previously by the authors.
The results are $M(\Xi_{cc}^{++})-M(\Xi_{cc}^+) = 1.41 \pm 0.12^{+0.76}$ MeV,
$M(\Xi_{bb}^0) - M(\Xi_{bb}^-) = - 4.78 \pm 0.06^{+0.03}$ MeV, and
 $M(\Xi_{bc}^+)-M(\Xi_{bc}^0)=-1.69 \pm0.07 ^{+0.39}$ MeV, where the statistical
errors reflect uncertainties in input mass splittings, and the systematic
errors are associated with the choice of constituent-quark masses.
\end{quote}
\smallskip

\leftline{PACS codes: 14.20.Lq, 14.20.Mr, 12.40.Yx}
\bigskip

%\draft

% This is Section I
\section{Introduction \label{sec:intro}}

Baryons with more than one heavy quark have proved to be elusive.  The SELEX
collaboration has presented evidence for several states \cite{Mattson:2002vu,%
Ocherashvili:2004hi,Engelfried:2007at}, but other experiments have not
confirmed them \cite{Ratti:2003ez,Chistov:2006zj,Aubert:2006qw,Aaij:2013voa,%
Kato:2013ynr}.  Simple constituent-quark models incorporating effective quark
masses, hyperfine interactions, and estimates of binding energies
\cite{Karliner:2008sv,Karliner:2006ny} have proved remarkably successful in
reproducing the masses of known hadrons with accuracies of several MeV.  In
agreement with most other estimates \cite{DeRujula:1975ge,bj,Anikeev:2001rk,%
Fleck:1989mb,Richard:1994ae,Korner:1994nh,Roncaglia:1995az,Lichtenberg:1995kg,%
Ebert:1996ec,SilvestreBrac:1996wp,Tong:1999qs,Gerasyuta:1999pc,Itoh:2000um,%
Kiselev:2001fw,Narodetskii:2002ib,Ebert:2002ig,Vijande:2004at,He:2004px,%
Richard:2005jz,Migura:2006ep,Albertus:2006ya,Roberts:2007ni,Gerasyuta:2008zy,%
Weng:2010rb,Zhang:2008rt} including ones using lattice gauge theory
\cite{Lewis:2001iz,Flynn:2003vz,Na:2008hz,Liu:2009jc,Namekawa:2012mp,%
Alexandrou:2012xk,Briceno:2012wt,Alexandrou:2014sha,Brown:2014ena}, this method
\cite{Karliner:2014gca} gives masses of $ccq$ ($q=u,d$) about 100 MeV above
the SELEX values, and close to the most recent lattice estimates
\cite{Brown:2014ena}.

The capability of the LHCb experiment to identify hadrons containing heavy
quarks makes it a prime instrument for determining the masses of the lowest
$\Xi_{cc}^{++} = ccu$ and $\Xi_{cc}^+ = ccd$ states.  As a benchmark, Ref.\
\cite{Karliner:2014gca} predicts $M(\Xi_{cc}) = 3627 \pm 12$ MeV for their
isospin average.  Their isospin splitting is then of interest, both as a
theoretical question and as a guide to further observation.  In particular,
the SELEX Collaboration reports large splittings whose values depend on
which of several bumps are assigned to the lowest $\Xi_{cc}$ states
\cite{Brodsky:2011zs}.  In the present paper we apply some simple methods,
used with previous success, to estimate isospin splittings in
the ground-state $\Xi_{cc}$, $\Xi_{bb}$, and $\Xi_{bc}$ baryons.  We describe
the methods in Sec.\ \ref{sec:meth}, present an alternative set of input
parameters in Sec.\ \ref{sec:alt}, quote results in Sec.\ \ref{sec:res},
and conclude in Sec.\ \ref{sec:conc}.

% This is Section II
\section{Methods \label{sec:meth}}

The impending improvement in the mass of the $\Xi^0$ baryon by the NA48
experiment at CERN \cite{Fanti:1999gy} and the KTeV experiment at Fermilab
led one of us \cite{Rosner:1998zc} to consider improved tests of relations
for baryon isomultiplet splittings.  A simple model was adopted which took
into account the intrinsic difference $\Delta= m_u - m_c$ between $u$ and $d$
quarks, Coulomb interactions $\Delta E_{ij~{\rm em}} = \alpha Q_i Q_j
\langle 1/r_{ij} \rangle$ between quarks, strong hyperfine (HF) interactions
\beq
\Delta E_{ij~{\rm HFs}} = {\rm const}\times \frac{|\Psi_{ij}(0)|^2 \langle
\sigma_i \cdot \sigma_j \rangle}{m_i m_j}~,
\eeq
and electromagnetic HF interactions
\beq
\Delta E_{ij~{\rm HFe}} = - \frac{2 \pi \alpha Q_i Q_j|\Psi_{ij}(0)|^2 \langle
\sigma_i \cdot \sigma_j \rangle}{3m_i m_j}~,
\eeq
where symbols are defined in Ref.\ \cite{Rosner:1998zc}.  We use the observed
mass splittings among the octet baryons \cite{PDG}, labeled with subscripts
denoting their $\Delta I$ values, summarized in Table \ref{tab:spl}, to
define the relative strengths of each contribution.

% This is Table I
\begin{table}
\caption{Experimental mass splittings between octet baryons \cite{PDG}.
\label{tab:spl}}
\begin{center}
\begin{tabular}{c c r} \hline \hline
Splitting & Symbol & Value (MeV) \\ \hline
$M(p) - M(n)$ & $N_1$ & --1.2933 \kern2.7em\\
$M(\Sigma^+) - M(\Sigma^-)$ & $\Sigma_1$ & $-8.08\kern0.45em 
\pm 0.08$\kern0.5em \\
$M(\Sigma^+) - 2M(\Sigma^0) + M(\Sigma^-)$ & $\Sigma_2$ & 
$1.535 \pm 0.090$ \\
$M(\Xi^0) - M(\Xi^-)$ & $\Xi_1$ & $-6.85 \kern0.45em\pm 0.21$\kern0.5em \\ \hline \hline
\end{tabular}
\end{center}
\end{table}

Each of these splittings may be expressed as a function of four unknowns
$\Delta$ (intrinsic $u-d$ mass difference), $a$ (Coulomb interaction), $b$
(strong HF interaction), and $c$ (electromagnetic HF interaction), where we
have simplified the notation of Ref.\ \cite{Rosner:1998zc} and neglected
effects of two-body kinetic energy operators:
\bea
N_1 & = & \Delta + \frac{a}{3} + b \left( \frac{1}{m_u^2} - \frac{1}{m_d^2}
\right) + \frac{c}{9} \left( \frac{4}{m_u^2} - \frac{1}{m_d^2} \right) \\
\Sigma_1 & = & N_1 + \Xi_1 \\
\Sigma_2 & \simeq & a + \frac{c}{{\bar m}^2} \\
\Xi_1 & = & \Delta - \frac{2a}{3} + b \left( \frac{4}{m_dm_s} -\frac{4}{m_um_s}
\right) + \frac{c}{9} \left( \frac{4}{m_d m_s} + \frac{8} {m_u m_s} \right)~,
\eea
where $\bar m$ is the average of $m_u$ and $m_d$, and we have neglected a
term of second order in $\Delta$ in $\Sigma_2$.  We have written a shorthand
for $\Sigma_1$ since under the present assumptions it satisfies the
Coleman-Glashow relation $\Sigma_1 = N_1 + \Xi_1$ 
\cite{CG} and is not independent.  Given quark masses
and an estimate of strong hyperfine structure from the splitting between the
$\Delta$ resonance and the nucleon (fixing $b$), one can determine the three
free parameters $\Delta$, $a$, and $\gamma \equiv c/\bar m^2$.  

Similar methods lead to estimates for isospin splittings in baryons with
two heavy quarks.  The results, after neglecting terms of second order in
$\Delta$, and defining $\beta \equiv b/\bar m^2$,
are
\bea
\Xi_{cc,1} \equiv M(\Xi_{cc}^{++}) - M(\Xi_{cc}^+) & = & \Delta + \frac{4a}{3}
+ \frac{4 \beta \Delta}{m_c} - \frac{8 \gamma \bar m}{3 m_c}~,\\
\Xi_{bb,1} \equiv M(\Xi_{bb}^0) - M(\Xi_{bb}^-) & = & \Delta - \frac{2a}{3}
+ \frac{4 \beta \Delta}{m_b} + \frac{4 \gamma \bar m}{3 m_b}~,\\
\Xi_{bc,1} \equiv M(\Xi_{bc}^+) - M(\Xi_{bc}^0) & = & \Delta + \frac{a}{3}
+ 2 \beta \Delta \left(\frac{1}{m_c} + \frac{1}{m_b} \right)
+ \frac{\gamma \bar m}{3} \left( \frac{2}{m_b} - \frac{4}{m_c} \right)~.
\eea

In order to specify $\Delta,~a,$ and $\gamma$ we must choose a set of
constituent-quark masses.  This was done in Ref.\ \cite{Karliner:2016zzc},
in two models, depending on whether or not a universal set of masses was
chosen for mesons and baryons.  In this section we shall consider quark
masses which fit both baryons and mesons simultaneously, with an added
``string-junction'' contribution $S = 161.5$ MeV for baryons.  Such an additive
constant does not affect mass {\it differences}, with which we are concerned
here.  (The alternative set is considered in the next section.)  Thus we take
$\bar m = 308.5$ MeV, $m_s = 482.2$ MeV, $\beta = 50.4$ MeV, $m_c=1655.6$ MeV,
and $m_b = 4988.6$ MeV.  A fit to octet baryon masses then give $\Delta =
- 2.681$ MeV, $a = 2.830$ MeV, $\gamma = -1.295$ MeV, and contributions
summarized in Table \ref{tab:sq}.  Here we have fixed $N_1$ at its measured
value of -1.2933 MeV, as its experimental error is negligible.
The uncertainties are those generated by varying each octet-baryon splitting
by $1 \sigma$ and adding the errors in quadrature.

% This is Table II
\begin{table}
\caption{Contributions to isospin splittings (MeV) using universal
constituent-quark masses in mesons and baryons.
\label{tab:sq}}
\begin{center}
\begin{tabular}{c r r r r r r r} \hline \hline
 & $N_1$\, & $\Sigma_1$\, & $\Sigma_2$\, & $\Xi_1$\, & $\Xi_{cc,1}$ &
$\Xi_{bb,1}$ &
  $\Xi_{bc,1}$ \\ \hline
$m_u - m_d$ & --2.68 & --5.36 & 0.00 & --2.68 & --2.68 & --2.68 & --2.68 \\
Coulomb & 0.94 & --0.94 & 2.83 & --1.89 & 3.77 & -1.89 & 0.94 \\
StrHF & 0.88 & -0.24 & 0.00 & -1.12 & -0.33 & --0.11 & --0.22 \\
EMHF & --0.43 & --1.54 & --1.30 & --1.11 & 0.64 & --0.11 & 0.27 \\
Total & --1.293\kern-0.4em & --8.086\kern-0.4em & 1.535\kern-0.4em &
-6.793\kern-0.4em & 1.409\kern-0.4em & --4.783\kern-0.4em &
-1.687\kern-0.4em \\
 & & & & & $\pm 0.116$\kern-0.4em& $\pm 0.058$\kern-0.4em & $\pm
0.067$\kern-0.4em \\
 \hline \hline
\end{tabular}
\end{center}
\end{table}

Note that the $\Delta I = 2$ mass difference is fitted {\it exactly}.  The
$\chi^2$ for this fit is 0.083, of which 0.010 comes from $\Sigma_1$
and 0.073 comes from $\Xi_1$.  This is just the extent to which the
Coleman-Glashow relation is obeyed.

\section{Alternative parameters \label{sec:alt}}

In a model in which mesons and baryons are described by separate
constituent-quark masses \cite{Karliner:2016zzc}, the parameters are
$\bar m = 363.7$ MeV, $m_s = 536.3$ MeV, $\beta = 49.3$ MeV, $m_c = 1710.5$
MeV, and $m_b = 5043.3$ MeV.  The fit gives $\Delta = -2.476$ MeV, $a = 3.053$
MeV, and $\gamma = -1.518$ MeV.  The results are shown in Table \ref{tab:bq}.
The uncertainties are those generated by varying each octet-baryon splitting
by $1 \sigma$ and adding the errors in quadrature.

% This is Table III
\begin{table}
\caption{Contributions to isospin splittings (MeV) using separate 
constituent-quark masses in mesons and baryons.
\label{tab:bq}}
\begin{center}
\begin{tabular}{l r r r r r r r} \hline \hline
 & $N_1$ & $\Sigma_1$ & $\Sigma_2$ & $\Xi_1$ & $\Xi_{cc,1}$ & $\Xi_{bb,1}$ &
  $\Xi_{bc,1}$ \\ \hline
$m_u-m_d$ & --2.48 & --4.95 & 0.00 & --2.48 & --2.48 & --2.48 & --2.48 \\
Coulomb & 1.02 & --1.02 & 3.05 & --2.04 & 4.07 & --2.04 & 1.02 \\
StrHF & 0.67 & --0.24 & 0.00 & --0.91 & --0.29 & --0.10 & --0.19 \\
EMHF & --0.51 & --1.88 & --1.52 & --1.37 & 0.86 & --0.15 & 0.36 \\
Total & --1.293\kern-0.5em & --8.086\kern-0.5em  & 1.535\kern-0.5em  &
--6.793\kern-0.5em  & 2.167\kern-0.5em  & --4.754\kern-0.5em  &
--1.293\kern-0.5em  \\
 & & & & & $\pm 0.109$\kern-0.5em & $\pm 0.058$\kern-0.5em 
& $\pm 0.062$\kern-0.5em \\
\hline \hline
\end{tabular}
\end{center}
\end{table}

The fit again reproduces the value of $\Sigma_2$ exactly, obtains the same
values for $\Sigma_1$ and $\Xi_1$, and thus has the same individual and overall
$\chi^2$ values.

\section{Results \label{sec:res}}

A slight preference for the string-based constituent-quark masses was
expressed in Ref.\ \cite{Karliner:2016zzc}.  Hence we shall quote predictions
for isospin splittings based on that model, with a systematic error associated
with the possible choice of independent constituent-quark masses for mesons
and baryons.  The results are:
$M(\Xi_{cc}^{++})-M(\Xi_{cc}^+) = 1.41 \pm 0.12^{+0.76}$ MeV,
$M(\Xi_{bb}^0) - M(\Xi_{bb}^-) = - 4.78 \pm 0.06^{+0.03}$ MeV, and
 $M(\Xi_{bc}^+) - M(\Xi_{bc}^0)= -1.69 \pm 0.07^{+0.39}$ MeV.
The first error is the greater of two very similar statistical errors in
Tables II and III.

Some approaches give values consistent with ours.  Ref.\ \cite{Hwang:2008dj}
finds $\Xi_{cc,1} = 2.3 \pm 1.7$ MeV, $\Xi_{bb,1} = - 5.3 \pm 1.1$ MeV, and
$\Xi_{bc,1} = - 1.5 \pm 0.9$ MeV.  Ref.\ \cite{Brodsky:2011zs} finds $1.5 \pm
2.7$ MeV, $- 6.3 \pm 1.7$ MeV, and $-0.9 \pm 1.8$ MeV for these quantities,
while a lattice-QCD-based approach \cite{Borsanyi:2014jba} finds $\Xi_{cc,1} =
(2.16)(11)(17)$ MeV, slightly favoring our set of independent quark masses for
mesons and baryons.  These results, along with some others, are compared in
Table IV.

% This is Table IV
\def\Astrut{\vrule height 2.4ex depth 1.0ex width 0pt}
\begin{table}
\caption{Comparison of predictions for isospin splittings (MeV)
in doubly heavy baryons. 
\label{tab:comp}}
\begin{center}
\begin{tabular}{c r r r} \hline \hline
\Astrut
  Reference 
& $\Xi_{cc,1}$ \kern2.0em
& $\Xi_{bb,1}$ \kern2.0em
& $\Xi_{bc,1}$ \kern2.0em 
\\ \hline\hline 
\Astrut This work 
& $1.41    \pm 0.12^{{+}0.76}$  
& \kern1em ${-}4.78 \pm 0.06^{{+}0.03}$
& ${-}1.69 \pm 0.07^{{+}0.39}$ 
\\ \hline
\Astrut\cite{Itoh:2000um}
& 4.7 \kern5.1em &&
\\ \hline
\Astrut \cite{Brodsky:2011zs} 
& $   1.5 \kern0.5em \pm 2.7$ \kern2.1em
& \kern1em ${-}6.3 \kern0.5em \pm 1.7$ \kern2.1em
& ${-}0.9 \kern0.5em \pm 1.8$ \kern2.1em
\\ \hline
\Astrut \cite{Hwang:2008dj}$^a$ \kern-0.7em
& $   2.3 \kern0.5em \pm 1.7$ \kern2.1em
& \kern1em ${-}5.3 \kern0.5em \pm 1.1$ \kern2.1em
& ${-}1.5 \kern0.5em \pm 0.9$ \kern2.1em
\\ \hline
\Astrut \cite{Borsanyi:2014jba}
& $2.16 \pm 0.11 \pm 0.12$ \kern-1.4em &&    
\\ \hline
\Astrut \cite{Lichtenberg:1977mv} 
& 4.7 \kern5.1em &&
\\ \hline
\Astrut \cite{Tiwari:1985ru}
& 1.11 \kern4.7em &&
\\ \hline
\Astrut \cite{Shah:2017liu}
& --9 \kern4.7em &&
\\ \hline \hline
\end{tabular} 
\end{center}
\vskip-1em
\strut\kern4.5em $^a$ 
{\small Ignores EM hyperfine interactions.}
\end{table}

\section{Discussion and conclusions\label{sec:conc}}

We have estimated isospin mass splittings in baryons $\Xi_{cc}$, $\Xi_{bb}$,
and $\Xi_{bc}$ containing two heavy quarks.  A major source of systematic
error, particularly in $\Xi_{cc,1} \equiv M(\Xi_{cc}^{++}) - M(\Xi_{cc}^{+})$,
is uncertainty in the choice of constituent-quark mass, giving $\Xi_{cc,1} =
1.41$ MeV for our favored model of universal quark masses in mesons and
baryons, while separate quark masses for mesons and baryons yield $\Xi_{cc,1} =
2.17$ MeV.

One assumption we have made concerns the universality of the expectation
value $\langle r_{ij} \rangle$ in evaluating the Coulomb self-energy.  It is
possible that two heavy quarks are more tightly bound to one another than
a light quark and a heavy one or two light quarks.  To lowest order, this
should not affect isospin splittings.  However, the difference between
binding of two light quarks from binding of a heavy quark with a light one
remains to be tested.  A start on this program was made in Sec.\ VI of
Ref.\ \cite{Rosner:1998zc}.  A relation $\Sigma_{c2} \equiv M(\Sigma_c^{++})
- 2 M(\Sigma_c^+) + M(\Sigma_c^+) = \Sigma_2$ was found there to be poorly
obeyed, but now reads $(1.92 \pm 0.82)$ MeV $ = (1.535 \pm 0.090)$ MeV, in
satisfactory agreement with the predicted equality.

It is worth recalling predicted lifetimes of baryons with two heavy quarks,
as the states with longer lifetimes are likely to be easier to distinguish
from background in a hadron collider.  Predictions by the authors are given
in Table XVI of Ref.\ \cite{Karliner:2014gca}, including $\tau(\Xi_{cc}^{++})
= 185$ fs and $\tau(\Xi_{cc}^{+}) = 53$ fs.  Most other predictions quoted
there are about three times as large, while preserving the ratio
$\tau(\Xi_{cc}^{++})/\tau(\Xi_{cc}^{+}) \simeq 3$.  The reason for the
shorter lifetime of $\Xi_{cc}^{+} = ccd$ is that the internal $W$ exchange
process $cd \to su$ is permitted, while it cannot occur for $\Xi_{cc}^{++} =
ccu$.  For a similar reason, one expects $\tau(\Xi_{bc}^+) > \tau(\Xi_{bc}^0)$
whereas $\tau(\Xi_{bb}^0) \simeq \tau(\Xi_{bb}^-)$.

We hope that these estimates
prove of use in discovery of such states.

\section*{Acknowledgements}
The work of J.L.R. was supported by the U.S. Department of Energy, Division of
High Energy Physics, Grant No.\ DE-FG02-13ER41958.

\end{document}